\input harvmac.tex
\vskip 2in
\Title{\vbox{\baselineskip12pt
\hbox to \hsize{\hfill}
\hbox to \hsize{\hfill IHES/97/P/xx}}}
{\vbox{\centerline{A Comment on a Fivebrane Term in Superalgebra}}}
\centerline{Dimitri Polyakov\footnote{$^\dagger$}
{polyakov@ihes.fr}}
\medskip
\centerline{\it Institut des Hautes Etudes Scientifiques}
\centerline{\it 35, Route de Chartres, F-91440 Bures-sur-Yvette, FRANCE}
\vskip .5in
\centerline {\bf Abstract}
We review our recent discussion of fivebrane central 
terms that appear in the 
 space-time superalgebra in $D=10$ provided that the 
space-time supercharges are taken in non-canonical pictures.
We correct the mistake contained in the original version
of the earlier paper 9703008 which suggested that the naive picture-changing
of the superalgebra gives rise to the non-perturbative five-form term.
This term vanishes because of gamma-matrix identity in $D=10$, as has
been pointed out by Berkovits in his recent paper.

\Date{April 97}
\vfill\eject

\lref\sez{E.Sezgin, hep-th/9609086, CTP-TAMU-27-96}
\lref\sezg{E.Sezgin, hep-th/9612220, CTP-TAMU-78-96}
\lref\shenker{D.Friedan,S.Shenker,E.Martinec,Nucl.Phys.{\bf B271}(1986) 93}
\lref\sezgi{P.S.Howe, E.Sezgin, P.C.West, KCL-TH-97-05, hep-th/9702008}
\lref\sagnotti{M.Bianchi, hep-th/9702093,ROM2F-97-4}
\lref\bianchi{A. Sagnotti, hep-th/9702098, ROM2F-97-6} 
\lref\sorokin{I.Bandos, K.Lechner, Nurmagambetov, P.Pasti, 
D.Sorokin, 
M.Tonin, hep-th/9701449 }
\lref\tseytlin{A.Tseytlin, Class.Quant.Grav.13:L81-L85 (1996)}
\lref\cvetic{M.Cvetic, A.Tseytlin, Phys.Rev. D53: 5619-5633, 1996}
\lref\polchino{J.Polchinski,NSF-ITP-95-122,hep-th/9510017}
\lref\townsend{P.Townsend,hep-th/9507048}
\lref\azc{J.A.de Azcarraga, J.P.Gauntlett, J.M.Izquierdo, P.K.Townsend,
 Phys.Rev.{\bf D63} (1989) 2443}
\lref\klebanov{S.Gubser,A.Hashimoto,I.R.Klebanov,J.Maldacena
Nucl.Phys.{\bf B472}(1996) 231}
\lref\green{M.B.Green, Phys.Lett.{\bf B223}(1989)157}
\lref\me{D.Polyakov, hep-th/9703008}
\lref\nb{N.Berkovits, hep-th/9704109}
\lref\olive{D.Olive, E.Witten, Phys.Lett.78B:97, 1978}
\lref\bern{J.N.Bernstein, D.A.Leites, Funkts. Analis i ego Pril.,
11 (1977)70-77}
\lref\cederwall{M.Cederwall, A.von Gussich,B.E.W.Nilsson,P.Sundell,
A.Westerberg, hep-th/9611159}
\lref\schwarz{M.Aganagic,J.Park, C.Popescu and J.H.Schwarz,
hep-th/9701166}
\lref\bel{A.Belopolsky, hep-th/9609220}
\lref\witten{E.Witten, hep-th/9610234}
\lref\witte{E.Witten, Nucl.Phys.{\bf B463} (1996), p.383}
\lref\myself{D.Polyakov, LANDAU-96-TMP-4, hep-th/9609092,
to appear in Nucl.Phys.{\bf B484}}
\lref\duff{M.J.Duff, K.S.Stelle, Phys.Lett.{\bf {B253}} (1991)113}
\lref\bars{I.Bars, hep-th/9607112}
\lref\grin{M.Green, Phys.Lett. {\bf B223}(1989)} 
\lref\banks{T.Banks, W.Fischler, S.Shenker, L.Susskind, hep-th/9610043}
\lref\seiberg{T.Banks, N.Seiberg, S.Shenker, RU-96-117, hep-th/9612157}
\lref\cohn{J.Cohn, D.Friedan, Z.Qiu, S.Shenker, Nucl.Phys.
{\bf B278}p.577(1986)}
In this letter, we review  the derivation of the five-form term,
in the ~\refs{\me}, showing that this term vanishes due to the
gamma-matrix identity in ten dimensions.We will argue, however,
that the fivebrane is present in the picture-changed superalgebra in an 
indirect way, as a total derivative three-form term.
This tree-form corresponds to a threebrane which is not fundamental, but 
rather is an intersection of two fivebranes.
  We start with recalling the operator product expansions
of the space-time spin operators with themselves and with the
worldsheet NSR fermions ~\refs{\cohn}.
We have:
\eqn\grav{\eqalign{\Sigma_\alpha (z) \Sigma^\beta (w) \sim 
{{{\delta_\alpha}^\beta}\over{(z-w)^{5/4}}}+
{{{{({\Gamma^m})}_\alpha}^\beta}\over{(z-w)^{3/4}}}\psi_m+
{{{{({\Gamma^{m_1m_2}})}_\alpha}^\beta}\over{(z-w)^{1/4}}}\psi_{m_1}\psi_{m_2}
+\cr+{\sum_p}
{{{{({\Gamma^{m_1...m_p}})}_\alpha}^\beta}\over{{(z-w)}^{5/4-p/2}}}
\psi_{m_1}...\psi_{m_5} + derivatives}}
Here the index m runs from 0 to 9, and $\Gamma^m$ are 32x32
ten-dimensional gamma-matrices.The fermionic indices $\alpha, \beta,...$
are alternated by the charge conjugation matrix
$C_{\alpha\beta}= (\Gamma^0)_{\alpha\beta}$.
The O.P.E. between $\Sigma$ and $\psi$ is given by:
~\refs{\shenker, \cohn}:
\eqn\lowen{\psi^m (z) \Sigma_\alpha (w) \sim 
{{{{(\Gamma^m)_\alpha}^\beta}\Sigma_\beta(w)}\over{(z-w)^{1/2}}}}
Alternatively, by raising the index $\alpha$ simultaneously
in the l.h.s and the r.h.s. of the last equation ( by multiplying
both the l.h.s. and the r.h.s. by $\Gamma^0$ from the left),
we can write the O.P.E. as
\eqn\lowen{\psi^m (z) \Sigma^\alpha (w) \sim 
{{{(\Gamma^m)^{\alpha\beta}\Sigma_\beta(w)}\over{{(z-w)^{1/2}}}}} =
{{{({\Gamma^m})^\alpha}_\beta{\Sigma^\beta}(w)}\over{(z-w)^{1/2}}}}
Next, the expression for the space-time supercharges in the canonical picture
is given by:
\eqn\lowen{Q_\alpha = \oint{{dz}\over{2i\pi}} e^{-1/2\phi}\Sigma_\alpha(z)}
where $\phi$ is bosonized superconformal ghost field ~\refs{\shenker}.
As is easy to check,
the computation of the anticommutator of two supercharges in this
canonical picture, by using the O.P.E. (1) gives the standard result:
\eqn\lowen{\lbrace{Q_\alpha, Q^\beta}\rbrace = {{(\Gamma^m)}_\alpha}^\beta
P_m},
where
$P_m = \oint {{dz}\over{2i\pi}} e^{-\phi}\psi_m(z)$ is the momentum
operator in the $-1$-picture.
Now, we would like to consider the same anticommutator 
with the supercharges $Q_\alpha$, $Q^\beta$ being taken in the 
non-canonical $+1/2$-picture. These supercharges are obtained from 
those of the formula (5) by the picture-changing transformation,
implemented by the picture-changing operator
\eqn\lowen{:\Gamma_1: = e^{\phi}\psi^m\partial{X_m}+ ghosts}
The ghost terms will be dropped in our calculation since they produce
the contributions not significant for the correlation functions.
By using the O.P.E's (2), (3), we find that the expressions for the 
supercharges in the $+1/2$-picture are given by:
\eqn\grav{\eqalign{Q_\alpha^{(+1/2)}= \oint{{dz}\over{2i\pi}}
e^{1/2\phi}{{{(\Gamma^m)}_\alpha}^\beta}\Sigma_\beta\partial{X_m}\cr
Q^{\alpha(+1/2)}=\oint{{dz}\over{2i\pi}}
e^{1/2\phi}{{{(\Gamma^m)}^\alpha}_\beta}{\Sigma^\beta}\partial{X_m}}}
The evaluation of the anticommutator, again by using the O.P.E. (1)
gives:
\eqn\grav{\eqalign{\lbrace{Q_\alpha}^{(+1/2)},{Q^\beta}^{(+1/2)}\rbrace=
{{({\Gamma^m})_\alpha}^\beta}P_m^{(+1)}+\cr+
{{(\Gamma^n)_\alpha}^\gamma}{{(\Gamma^{m_1...m_5})_\gamma}^\delta}
{{(\Gamma_n)^\beta}_{\delta}}Z_{m_1...m_5}}}
Here $P_m^{(+1)}$ is the momentum operator in the $+1$-picture
(the expression for it is given in ~\refs{\me}) and
the five-form central charge $Z$ is found in ~\refs{\me} to be 
equal to:
\eqn\lowen{Z_{m_1...m_5}=e^\phi{\psi_{m_1}...\psi_{m_5}}}
Let us analyze now the gamma-matrix factor 
in the  five-form term in the last equation.
 We have:
\eqn\grav{\eqalign{{{(\Gamma^n)_\alpha}^\gamma}
{{(\Gamma^{m_1...m_5})_\gamma}^\delta}{{(\Gamma_n)^\beta}_\delta}=
(\Gamma^n)_{\alpha\gamma}(\Gamma^{m_1...m_5})^{\gamma\delta}
{{(\Gamma^T_n)_\delta}^\beta}=\cr
=(\Gamma^n)_{\alpha\gamma}(\Gamma^{m_1...m_5})^{\gamma\delta}
(\Gamma_n^T)_{\delta\rho}{\Gamma_0}^{\rho\beta}=\cr
= 2{{(\Gamma^{m_1...m_5})_{\alpha}}^\beta}}}
But for the gamma-matrices with both indices down
$\Gamma^m = (\Gamma^m)^T$
Therefore the gamma-matrix factor in front of the five-form term
is proportional to $\Gamma^m \Gamma^{m_1...m_5}\Gamma_m$, and
this factor vanishes in $D=10$
This concludes our analysis of the gamma-matrix factor in the
fivebrane term.
The fivebrane term, as we see, does appear in the superalgebra with
the supercharges taken in the non-canonical picture.
 We conclude  by adding few observations about the space-time superalgebras
in non-canonical pictures .
While the five-form term vanishes in the anticommutator
$\lbrace{Q^{+{1\over2}},Q^{+{1\over2}}}\rbrace$,
 there is another p-form term, namely, the three-form  
that appears in the space-time $supercurrent$ algebra.
Being the total derivative, it is proportional to:
\eqn\grav{\eqalign{Z_{m_1...m_3} 
\sim \partial (e^\phi \psi_{m_1}...\psi_{m_3})}}
It is known that two fivebranes intersect over a threebrane .
The total derivative seems to be related to the fact that the
threebrane is not fundamental, but rather is an intersection of 
two fivebranes. Therefore, in an indirect way, the fivebrane does
appear in the picture-changed superalgebra.
Therefore the information about the dynamics of intersecting
branes  may be hidden in the concrete structure of space-time
supercurrent algebra.
I'm grateful to N.Berkovits and E.Witten for pointing out to me
the error in the original version of my paper hep-th/9703008,
in which the vanishing $\Gamma$-matrix factor was overlooked. 
The argument about the vanishing if the fivebrane term has 
been given earlier in ~\refs{\nb}
\listrefs
\end